\documentclass[runningheads]{svmult}

\usepackage{graphicx}  
\usepackage{physprbb}  

\newcommand{\greeksym}[1]{{\usefont{U}{psy}{m}{n}#1}}
\newcommand{\umu}{\mbox{\greeksym{m}}}

\newcommand{\equ}[1]{eq.~(\ref{eq:#1})}

\newcommand{\se}[1]{\S\ref{sec:#1}}
\newcommand{\fig}[1]{Fig.~\ref{fig:#1}}
\newcommand{\be}{\begin{equation}}
\newcommand{\ee}{\end{equation}}

\newcommand{\msun}{M_\odot}
\newcommand{\ifm}[1]{\relax\ifmmode#1\else$\mathsurround=0pt #1$\fi}
\newcommand{\kms}{\ifmmode\,{\rm km}\,{\rm s}^{-1}\else km$\,$s$^{-1}$\fi}

\newcommand{\hkpc}{\,\ifm{h^{-1}}{\rm kpc}}
\newcommand{\ltsima}{$\; \buildrel < \over \sim \;$}
\newcommand{\lsim}{\lower.5ex\hbox{\ltsima}}
\newcommand{\gtsima}{$\; \buildrel > \over \sim \;$}
\newcommand{\gsim}{\lower.5ex\hbox{\gtsima}}
\newcommand{\prop}{\propto}

\newcommand{\Ms}{M_*}
\newcommand{\gamef}{\gamma_{\rm eff}}
\newcommand{\gamc}{\gamma_{\rm crit}}

\begin{document}
\title*{Characteristic Scales in Galaxy Formation}
%
%
\toctitle{Characteristic Scales in Galaxy Formation}
\titlerunning{Characteristic Scales in Galaxy Formation}
\author{Avishai Dekel\inst{1,2}}
%
\authorrunning{Avishai Dekel}
%
%
\institute{The Hebrew University of Jerusalem, Israel
\and Institut d'Astrophysique and Observatoire de Paris, France}

\maketitle              

\begin{abstract}
Recent data, e.g. from SDSS and 2dF, 
reveal a robust \emph{bi-modality} in the distribution
of galaxy properties, with a characteristic transition scale at stellar mass
$\Ms \sim 3\times 10^{10}\msun$ (near $L_*$), 
corresponding to virial velocity $V\sim 100\kms$.
Smaller galaxies tend to be blue disks of young populations. They define
a ``fundamental line" of decreasing surface brightness, metallicity and
velocity with decreasing $\Ms$, which extends to the smallest dwarf
galaxies.  Galaxies above the critical scale are dominated by red
spheroids of old populations, with roughly constant high surface
brightens and metallicity, and they tend to host AGNs. 
A minimum in the virial M/L is obtained at the same magic scale.
This bi-modality can be the combined imprint of several different 
physical processes.
On smaller scales, disks are built by \emph{cold flows}, and 
\emph{supernova feedback} is effective in regulating star formation. 
On larger scales, the infalling gas is \emph{heated by a virial shock}
and star formation can be suppressed by \emph{AGN feedback}.
Another feedback mechanism -- gas evaporation due to photo-ionization -- may
explain the existence of totally dark halos below $V\sim 30\kms$. 
The standard cooling barriers are responsible for the loose upper and 
lower bounds for galaxies: $10<V<300\kms$.
\end{abstract}

\section{Introduction}
\label{sec:intro}

\emph{Observations} indicate four characteristic scales in the global properties
of galaxies, expressed 
either in terms of virial velocity $V$, or stellar mass $\Ms$:
\begin{itemize}
\item
An upper limit at $V\sim {\bf 300}\kms$, which is the lower bound for clusters.
\item
A robust bi-modality scale at $\Ms \simeq 3\times 10^{10}\msun$ \cite{kauf03}, 
corresponding to $V\sim {\bf 100}\kms$. 
This is comparable to $L_*$ of the Schechter luminosity function, marking
the exponential upper bound for disk galaxies.
Smaller galaxies tend to be blue, star-forming disks with
the surface brightness and metallicity decreasing with decreasing $\Ms$ (LSB).
More massive galaxies are dominated by red, old-population spheroids with
high surface brightness and metallicity insensitive to $\Ms$ (HSB)
\cite{kauf03}. AGNs typically reside in halos above the bi-modality scale
\cite{kauf04}.
The halo M/L shows a minimum at the same scale \cite{yang03}.
\item
Dark-dark halos (DDH) with no luminous components seem to dominate the
halo distribution below  $V\sim {\bf 30}\kms$. 
The relatively flat faint-end luminosity function and the steeper mass function
predicted by the standard $\Lambda$CDM cosmology cannot be reconciled with 
the virial theorem and the Tully-Fisher relation obeyed by the luminous
galaxies (\se{evap}). Also,
most dwarf spheroidals in the Local Group
lie in this range, while dwarf irregulars are typically $V>30\kms$
\cite{mateo98} \cite{vdbergh00} \cite{grebel03} 
\cite{dw03} \cite{wdc04}.
\item
A lower bound at $V\sim {\bf 10}\kms$ for dwarf galaxies (see \cite{dw03}).
\end{itemize}

We summarize here how \emph{theory} predicts the following characteristic 
scales:
\begin{itemize}
\item
A loose upper bound for cooling (by bremsstrahlung) 
on a dynamical time at $M\sim 10^{12-13}\msun$, 
corresponding to $V\sim 300\kms$.
\item
A virial shock-heating scale at {\bf $\Ms \sim 3\times 10^{10}\msun$}, 
or $V\sim 100\kms$.
In less massive halos virial shocks cannot form and the gas flows cold
into the disk vicinity (\se{shock}).
\item
A supernova-feedback scale at {\bf $V\sim 100\kms$}, below which the potential
wells are shallow enough for the energy fed to the ISM by supernova to
significantly heat the halo gas, or even blow it out (\se{supernova}).
\item
A photo-ionization scale at $V\sim 30\kms$. Once the gas is continuously
ionized by the background UV flux, IGM gas cannot accrete into smaller halos
and gas already in such halos evaporates via a steady thermal wind
(\se{evap}).
\item
A lower bound for cooling (by atomic hydrogen) at $V\sim 10\kms$.
\end{itemize}

We briefly address the origin of these scales, 
and try to associate them with the observed scales.

\section{Shock-Heating Scale}
\label{sec:shock}

The standard picture of disk formation is that after the dark halo
virializes, the infalling gas is shock heated near the virial radius. 
It then cools radiatively and, if the cooling time is shorter than the
dynamical time, the gas accretes onto a central disk and forms stars with 
the associated feedback. The upper-limit halo mass for efficient 
cooling is $M_{\rm cool} \sim 10^{12-13}\msun$ 
\cite{ro77} \cite{wr78}.
The common wisdom used to be that this explains the upper bound for
disk galaxies, but it has been realized that the predicted
scale is somewhat higher than the observed scale.

\begin{figure}
\begin{center} 
\hspace{-7pt}
\begin{minipage}{0.5\textwidth} \centering
\includegraphics[width=1.0\textwidth]
{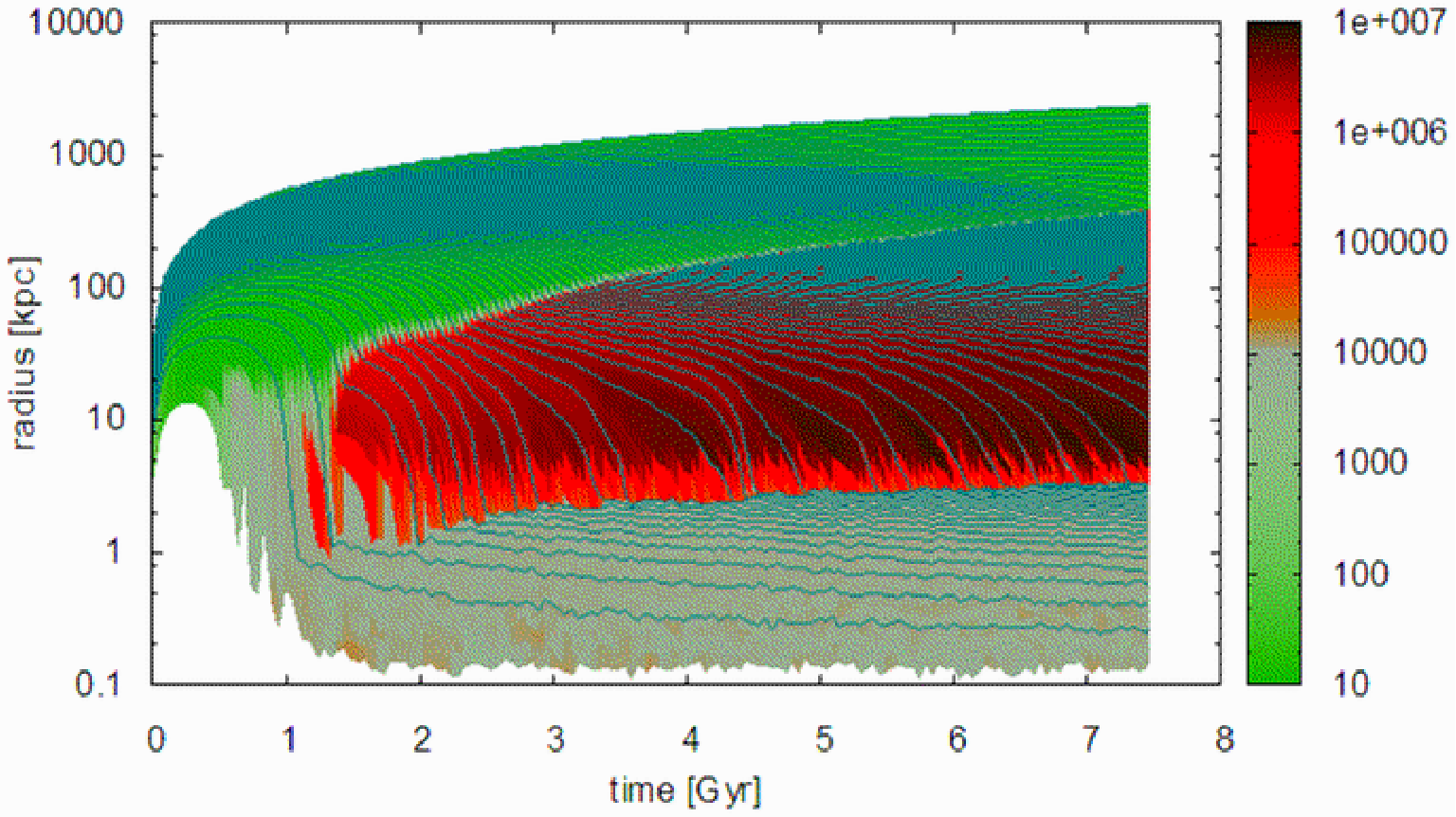}
\end{minipage}%
\hspace{0pt}
\hfill
\begin{minipage}{0.5\textwidth}
\centering
\includegraphics[width=1.0\textwidth]
{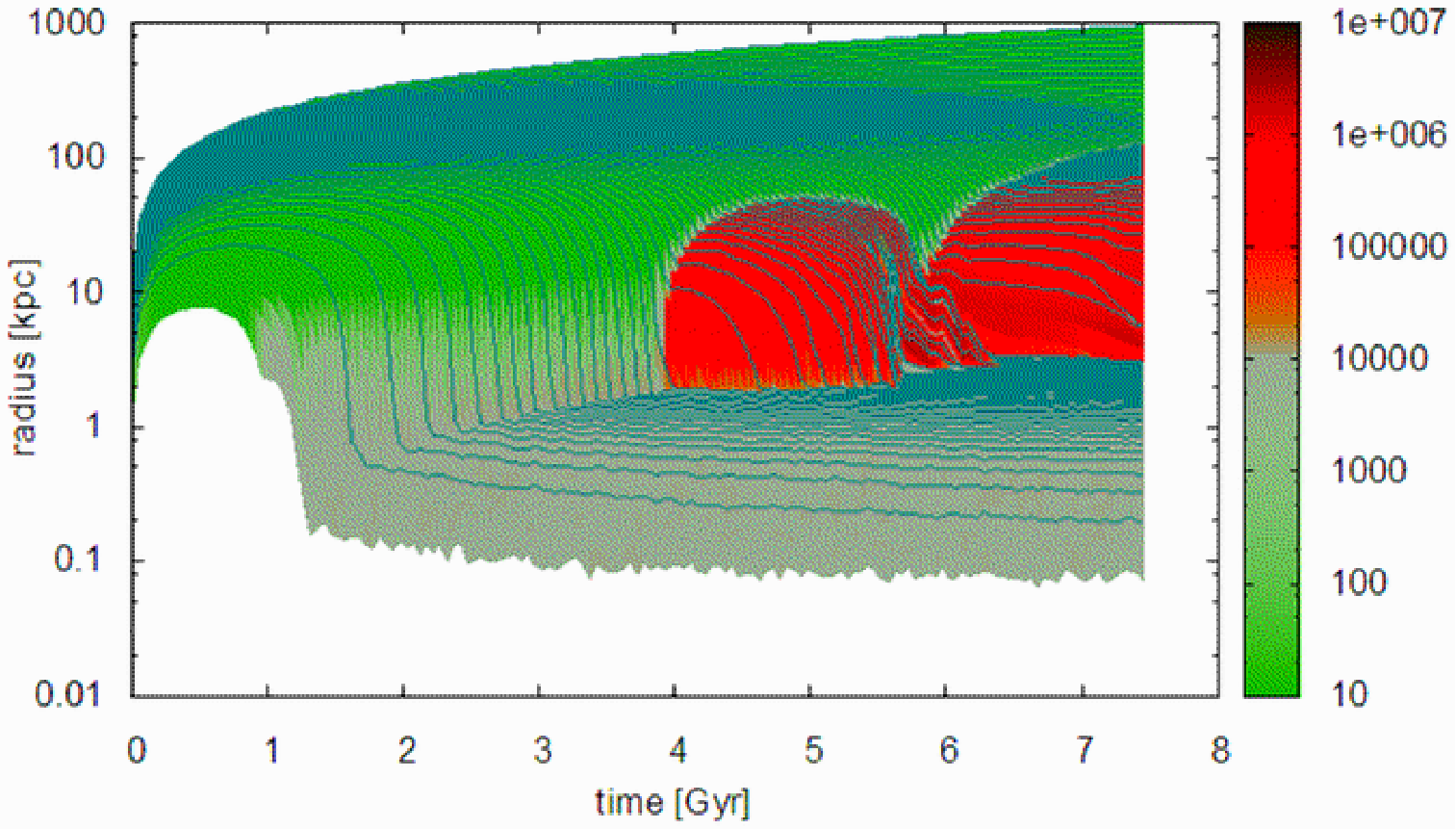}
\end{minipage} \hspace*{\fill}
\end{center}
\vspace{-10pt}
\caption[]{Time evolution of the radii of Lagrangian
gas shells (lines) in a spherical simulation of a protogalaxy
consisting of primordial gas and dark matter.
Temperature is marked by color.
A shock shows up as a sharp break in the flow lines and as an
abrupt change in temperature.
The lower discontinuity marks the ``disk" radius, formed due
to an artificial centrifugal force.
(\textbf{a}) A massive system, where the virialized masses
are in the range $10^{11}-10^{13}\msun$. 
(\textbf{a}) A less massive system, $10^{10}-10^{12}\msun$.
A virial shock exists only in systems more massive than $>10^{11}\msun$.
} 
\label{fig:bd03}
\end{figure}

\fig{bd03} shows the time evolution of the radii of Lagrangian gas shells 
in a spherical system consisting of gas (with primordial composition)
and dark matter, as simulated by Birnboim \& Dekel \cite{bd03} 
using a 1D hydro code.
Not shown are the dark-matter shells, which collapse, oscillate and
tend to increase the gravitational attraction exerted on the gas shells.
The left panel focuses on massive halos of $\sim 10^{12}\msun$.
Indeed, a shock exists near the virial radius at all times;
it gradually propagates outward, encompassing more mass in 
time.  The hot post-shock gas is in a quasi-static equilibrium,
pressure supported against gravity.
The right panel focuses on halo masses smaller by an order of magnitude.
An interesting new phenomenon is seen here, where
a stable shock forms and propagates from the disk toward the virial radius
only after a mass of $\sim 10^{11}\msun$ has virialized.
In less massive systems, the cooling rate is faster than the infall
time, making the ``post-shock" gas unstable against gravitational infall
and thus not being able to support an extended shock.

This behavior can be understood via the following instability analysis 
of the gas just behind the shock \cite{bd03}.
The standard equation of state is written as
$P=(\gamma-1)\rho\, e$
where $\rho$ and $e$ are the gas density and internal energy,
and $\gamma =5/3$ for a mono-atomic gas.
In the no-cooling case, the adiabatic index is defined as
$\gamma \equiv (\partial \ln P/\partial \ln \rho)_{\rm ad}$,
and the system is known to be gravitationally stable once $\gamma>4/3$.
When there is radiative cooling at a rate $q$,
we define a new quantity:
\be
\gamef \equiv \left( {d \ln P \over d \ln \rho} \right)
= \gamma - {\rho \over \dot\rho}{q \over e} 
= \gamma - {t_{\rm dyn} \over t_{\rm cool}} \ .
\ee
The second equality follows from energy conservation, $\dot e = -P \dot{V} -q$,
plugged into the equation of state.
Note that $\gamef=\gamma$ when $q=0$,
and that the difference between the two is nothing but the ratio of
characteristic times
$t_{\rm dyn}/t_{\rm cool}$, where 
$t_{\rm dyn}\equiv \rho/\dot{\rho} \simeq r_{\rm s}/u$,
with $r_{\rm s}$ the shock radius and $u$ the infall velocity into it.
Using the jump conditions across the shock, 
one can express $\gamef$ in terms of the pre-shock gas quantities:
\be
\gamef = \gamma -{(\gamma+1)^4 \over 6(\gamma-1)^2}
  {\rho r_{\rm s} \Lambda(T) \over \left| u^3 \right|}, \quad
T={\mu \over k_{\rm B} N_{\rm A}}  {2\gamma -1 \over (\gamma+1)^2}  u^2 \ ,
\label{eq:gamef}
\ee
where $\rho$ is the pre-shock density. 

In order to test for stability, we perturb a shell by 
$r \rightarrow r+\delta r$ 
and compute the sign of the restoring force: $\ddot{\delta r} / \delta r$.
We assume a homologous infall $\delta r = u \delta t$ (supported by the
finding in the simulations above), and balancing forces
$\ddot r = -\rho^{-1} \nabla P - GM/r^3 =0$.
This leads to
\be
\ddot{\delta r} / \delta r \propto
\gamma -4/3 -2\gamef^{-1}(\gamma-\gamef)\ .
\ee
The system is stable when this expression is positive, namely for
\be
\gamef > \gamc = {2\gamma \over \gamma +2/3},
\quad {\rm or}\quad
{t_{\rm cool} \over t_{\rm dyn}} > {\gamma +2/3 \over \gamma^2-4\gamma/3} \ .
\label{eq:criterion}
\ee
For $\gamma=5/3$, we get $\gamc=10/7$ replacing the $\gamc=4/3$ of the
adiabatic case. The critical scale is then characterized by 
$t_{\rm cool}/t_{\rm dyn} = 21/5$ at the disk vicinity,
which translates to $t_{\rm cool}/t_{\rm dyn} \simeq 1$ at the virial radius.
Note that in our notation $t_{\rm dyn} \simeq R/V$ at the virial radius,
while the dynamical time used in the order-of-magnitude arguments of 
\cite{ro77} and \cite{wr78} is several times larger.

When $\gamef$ is computed using \equ{gamef}
just above the disk radius at every time of the spherical
simulation, we find that the criterion of \equ{criterion} indeed works
very well. Before the shock forms, it is well below $\gamc$ and gradually
rising, reaching $\gamc$ almost exactly when the shock starts propagating
outward. It then oscillates about $\gamc$ with a decreasing amplitude,
following the oscillations in the shock radius seen in \fig{bd03}. 

When put in a cosmological setting, we find that the critical halo mass for
virial-shock heating is
\be
M_{\rm shock} \simeq (1-6)\times 10^{11}\msun \, .
\ee
The higher estimate is obtained for near solar metallicity. 
This scale is comparable to the observed bi-modality scale,
and is quite insensitive to redshift. To see this, we use the 
spherical-collapse virial relations at $z$:
$M \propto (1+z)^{-3/2}V^3 \propto (1+z)^{3}R^3$,
and the general trend in the relevant range $\Lambda \prop T^{-1/2}$,
to obtain $t_{\rm cool} /t_{\rm dyn} \propto M$ independent of $z$.
Since at $z>2$
most halos are of $M<M_{\rm shock}$, we conclude that in most forming disks
the gas never heats to the virial temperature -- it flows cold all the way
to the disk vicinity. 

Although we are back to a condition involving $t_{\rm cool}/t_{\rm dyn}$,
which seems to remind us of the original criterion by Rees \& Ostriker,
it is now based on analyzing the actual physical process that could be
responsible for the heating to the virial temperature, namely the existance
of a virial shock. Because the dynamical time involved in the current analysis
is smaller than the one used in the original 
estimate by a factor of a few, our derived upper bound for halo masses
in which cold flows prevail is smaller by a similar factor.

\begin{figure}
\begin{center}
 \hfill \begin{minipage}{0.5\textwidth} \centering
 \includegraphics[width=5cm]
{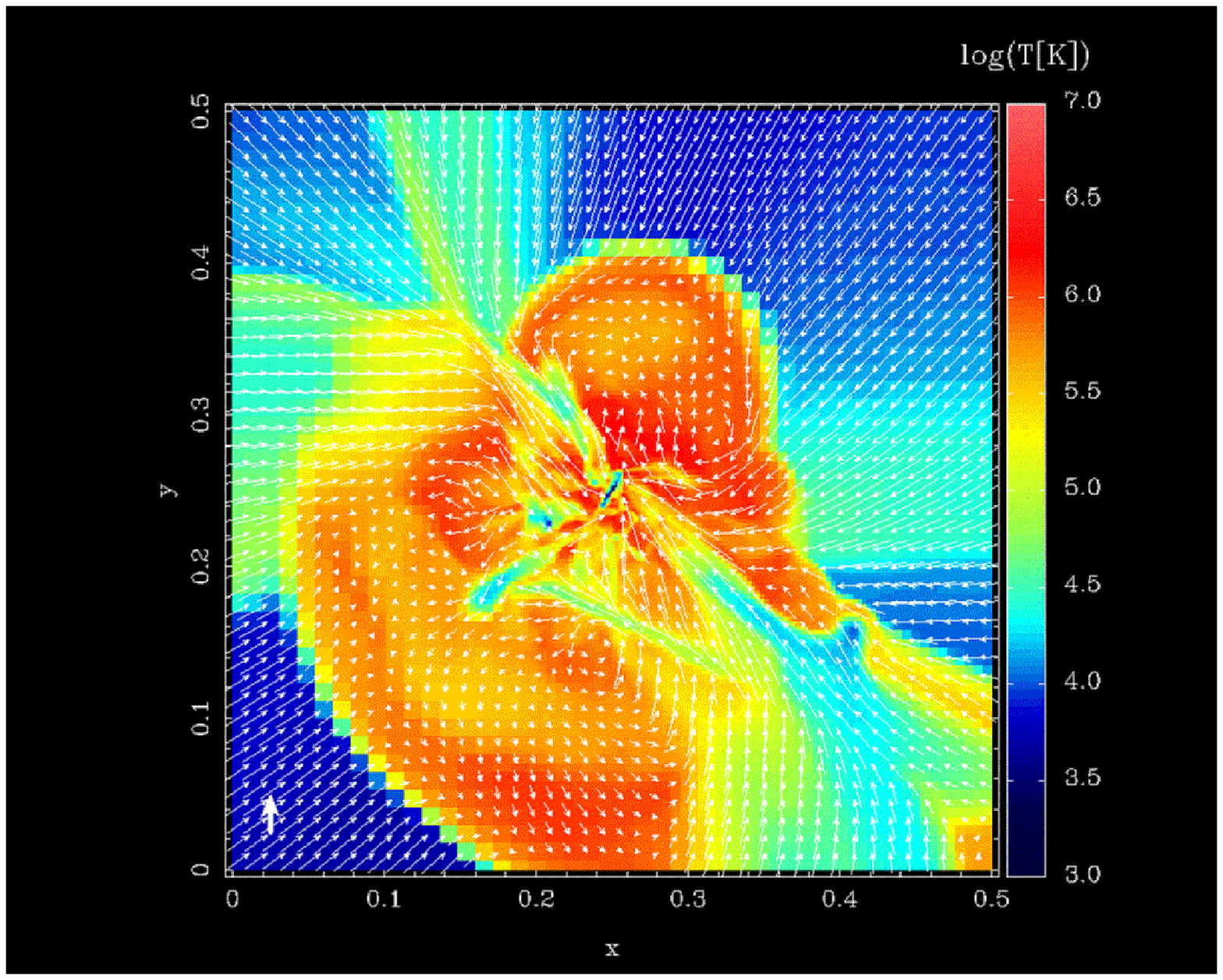} 
\end{minipage}
 \begin{minipage}{0.5\textwidth} \vspace{0.1cm} \centering
 \includegraphics[width=4.5cm]
{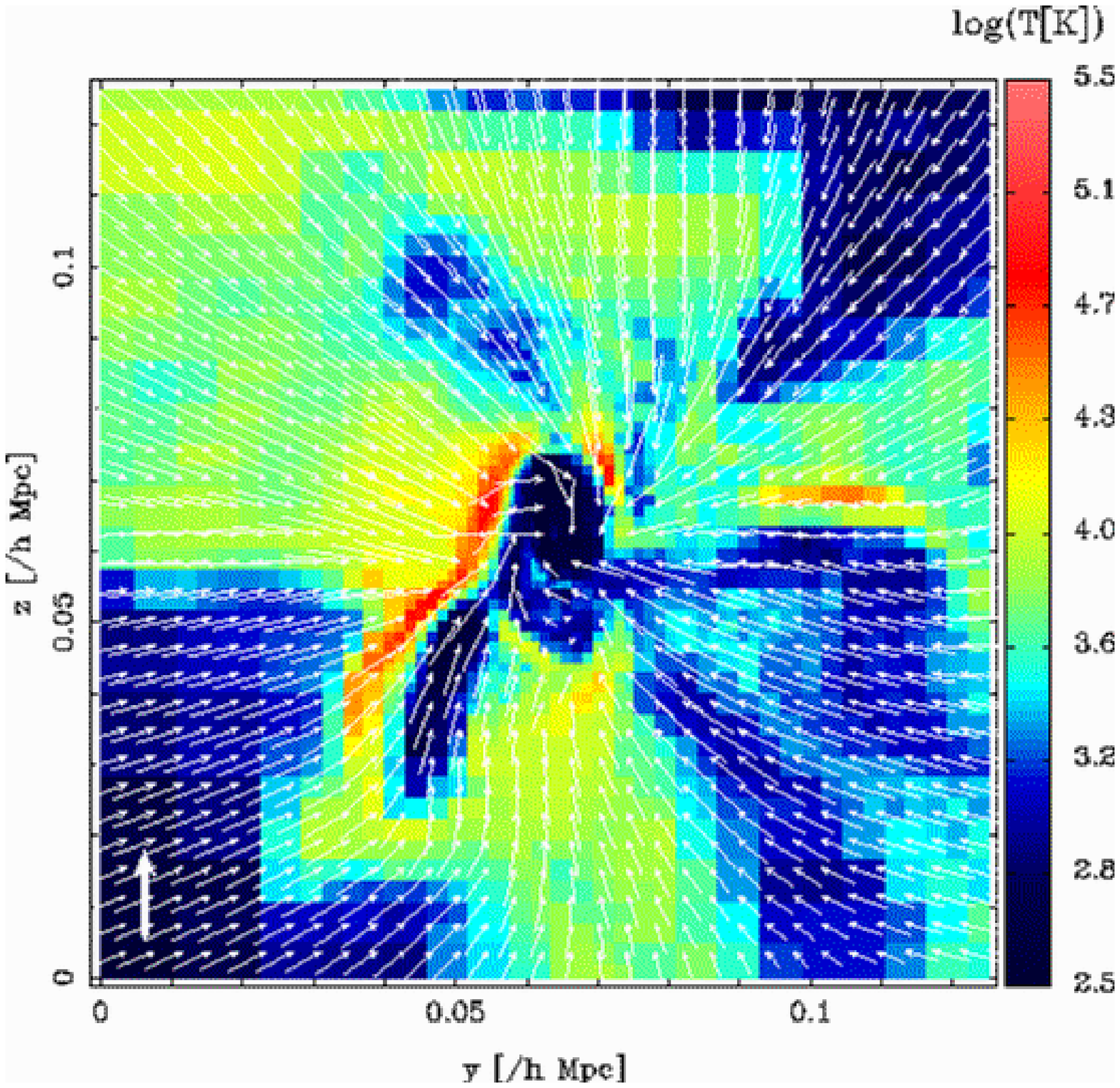}
\end{minipage}
\hspace*{\fill}
\end{center}
\vspace{-10pt}
\caption[]{Snapshots from cosmological hydro simulations (courtesy of Andrey
Kravtsov),
showing the gas temperature and flow fields in two protogalaxies.
(\textbf{a}) A halo of $M \simeq 3\times 10^{11}\msun$ at $z\simeq 4$,
showing a virial shock and heated gas penetrated by cold streams.
(\textbf{b}) A halo of $M \simeq 2\times 10^{10}\msun$ at $z\simeq 9$,
showing cold infall with little shock heating inside the virial radius
of $\sim 7 \hkpc$.}
\label{fig:krav} 
\end{figure}

Preliminary results from cosmological hydro simulations indicate that
the phenomenon is not restricted to spherical systems, and that our
stability criterion may in fact be approximately valid in the general
case where the inflow is along filaments.
\fig{krav} shows snapshots from Eulerian simulations by A. Kravtsov 
\cite{krav03} \cite{krav04},
showing the gas temperature and flow fields in two protogalaxies:
one of $M \simeq 3\times 10^{11}\msun$ at $z\simeq 4$,
and the other of $M \simeq 2\times 10^{10}\msun$ at $z\simeq 9$.
While the more massive galaxy shows a hot gas component at the virial 
temperature behind a virial shock, the smaller galaxy has mostly cold 
gas inside the virial radius.  
Also seen in the massive galaxy are cold streams penetrating the hot medium
toward the central disk. 
Similar results have been obtained by Fradal et al. \cite{fradal01}, 
who emphasis the feeding of galaxies by cold flows preferentially at 
early epochs, corresponding to less massive galaxies.
We are in a process of testing our shock-stability criterion using 
cosmological simulations, and generalizing it to the realistic case
of elongated streams.

Beyond introducing a new scale in galaxy formation,
the cold streaming toward the disk, 
as opposed to the commonly assumed shock heating and slow accretion,
may have several interesting implications. For example: 
\begin{itemize}
\item
It may modify the star formation rate and the associated supernova feedback.
If the result of the cold stream hitting the disk is a burst of stars
at high $z$,
this may help the semi-analytic models (SAMs) explain better the presence of
luminous red galaxies at $z\sim 1$.
This process should be properly incorporated in the SAMs.
\item
It may reduce the soft X-ray flux from halos. This is consistent 
with observed individual galaxies \cite{cioti91}, 
and may explain why the observed X-ray background is lower than the
predictions based on shock heating in all halos \cite{pen99}.
\item
The cold gas may emit a significant fraction of its infall energy
in Ly-$\alpha$, either if the streams cool and merge into the rotating disk
gradually without shocking \cite{keres04}, 
or after shocking at the disk vicinity. 
In the latter, the X-rays produced are likely to be captured inside a dense
Strongrem sphere of a few kpc and the energy may eventually be released 
by Ly-$\alpha$. 
This may explain the observed Ly-$\alpha$ emitters \cite{manning00}.
\end{itemize}

\section{Supernova Feedback Scale}
\label{sec:supernova}

Supernova feedback, which is a very different physical process, gives
rise to a characteristic scale in the same ballpark.
The energy fed to the ISM by supernovae can be written as
\be
E_{\rm SN} \simeq \nu \epsilon \dot{\Ms} t_{\rm rad}
\propto \Ms (t_{\rm cool}/t_{\rm dyn}) \ ,
\ee
where
$\nu$ is the number of supernovae per unit mass of forming stars,
$\epsilon$ is the typical supernova energy,
and $t_{\rm cool}$ is the time available for the supernova remnant to
share its energy with the medium during its adiabatic phase,
before a significant fraction of it
is radiated away. The second proportionality follows from the crude assumption
$\dot{\Ms} \sim \Ms/t_{\rm dyn}$.
Dekel \& Silk \cite{ds86} pointed out that for halos in the relevant
range, where the virial temperature is $T \sim 10^5$K, the cooling rate behaves 
roughly like $\Lambda \propto T^{-1}$, and then 
$t_{\rm cool}/t_{\rm dyn} \simeq 0.01$ for all galaxies.
This means that, despite the significant radiative losses, the energy 
fed to the ISM is 
$ E_{\rm SN} \prop \Ms$. 
When this energy input is compared to the energy required for significantly
heating the gas or blowing it out, 
$E_{\rm binding} \simeq M_{\rm gas} V^2$,
one ends up with the critical scale for effective feedback, 
\be
V_{\rm SN} \simeq 100\kms \, .
\ee
In shallower potential wells, the feedback from
a burst of stars can significantly suppress further star formation,
and produce low-surface-brightness (LSB) and dwarf galaxies.
We note that $V_{\rm SN}$ 
corresponds to $\Ms \simeq 3\times 10^{10}\msun$, comparable
to the observed bi-modality scale.

Tight correlations between the properties of galaxies below 
$3\times 10^{10}\msun$ define a ``\emph{fundamental line}" (FL)  
which stretches across five decades in luminosity down to the smallest dwarfs.
The mean scaling relations, involving $\Ms$, $V$, the metallicity $Z$ and the
surface brightness $\mu$, are approximately \cite{kauf03} \cite{dw03}
\cite{wdc04} \cite{tremonti04} 
\be
V\propto \Ms^{0.2}, \quad
Z\propto \Ms^{0.4}, \quad
\mu \propto \Ms^{0.6} \ .
\label{eq:scaling}
\ee
Note that the familiar Tully-Fisher relation at the bright end,  
$V\propto \Ms^{0.3}$ \cite{cour04}, seems to flatten in the dwarf regime.
Supernova feedback can explain the origin of the FL in simple terms
\cite{dw03}.
The above energy criterion, $E_{\rm SN} \propto \Ms \propto M_{\rm gas}V^2$,
with the assumption that the relevant gas mass is a constant fraction of the
halo mass, $M_{\rm gas} \propto M$, immediately implies the relation
driving the FL:
\be
\Ms/M \propto V^2 \ .
\label{eq:fl}
\ee
Assuming spherical collapse to virial equilibrium, the halo virial quantities
are related via,
$ M \propto V^3 \propto R^3$.
In the instantaneous recycling approximation, we
approximate $Z \prop \Ms / M_{\rm gas}$.  For the stellar radius we adopt
the standard assumption of angular-momentum conservation \cite{fe80}
\cite{mmw98},
$R_* \propto \lambda R$, with $\lambda$ a constant spin parameter, and 
use it in $\mu \propto \Ms /R_*^2$.
Combining the relations from \equ{fl} and on, we recover the scaling relations,
\equ{scaling}.
This success of this simplest possible toy model indicates that supernova
feedback could indeed be the primary driver of the FL below the bi-modality
scale, and it strengths the association of this process with the
origin of the transition scale itself.

\section{The Origin of Bi-Modality}
\label{sec:bimodality}

\begin{figure}
\begin{center}
\includegraphics[width=.8\textwidth]
{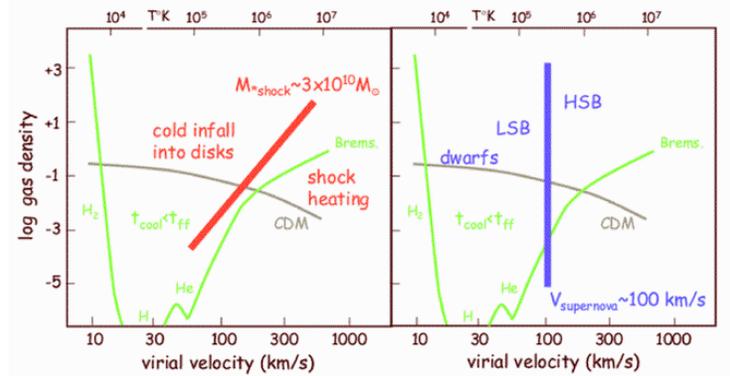}
\end{center}
\vspace{-10pt}
\caption[]{the characteristic scales in the classic diagram of gas density 
versus virial velocity. The horizontal (gray) curve marks 2-$\sigma$
halos in $\Lambda$CDM. The V-shaped (green) curve refers to 
$t_{\rm cool}=t_{\rm dyn}$; galaxies can form only above this line. 
(\textbf{a}) The shock-heating scale.
(\textbf{b}) The supernova feedback scale.
The two scales are comparable, 
at $V\sim 100\kms$ or $\Ms \sim 3\times 10^{10}\msun$.
}
\label{fig:cool}
\end{figure}

In \fig{cool}, we schematically put together the physical scales discussed
so far in the classic density-temperature diagram. 
Galaxies are loosely confined
to the region where $t_{\rm cool} < t_{\rm ff}$. 
The line $M_{*,{\rm shock}}\sim 3\times 10^{10}\msun$ 
distinguishes between massive galaxies which
suffer virial shock heating and lower mass galaxies where the gas flows
cold toward the disk. The line $V \sim 100\kms$ marks the transition 
from deep potential wells where HSBs form to shallow potential wells
where supernova feedback is effective and leads to LSBs and dwarfs along
the FL.

The two different physical processes happen to yield a similar characteristic
scale, and they thus combine to produce the bi-modality. A possible scenario is
as follows:
\begin{itemize}
\item
Halos below $M_{\rm shock}$ had cold infall to disks, possibly 
associated with bursts of star formation.
Galaxies above this threshold are built from progenitors that were 
below the threshold at high $z$ and thus formed early disks, which later
merged to spheroids of red, old stars.
Any halo gas in such massive halos consists of a hot medium and possibly 
cold streams.
\item
In halos below $V_{\rm SN}$, the SN feedback regulates the star-formation rate,
which leads to blue, young stars and the FL, $\Ms/M\propto V^2$, with $M/L$
rising toward smaller halos. The SN feedback may also suppress any AGN
activity in these halos.
\item
In more massive halos, the gas, which has been heated by virial shocks,
may be prevented from cooling by AGN feedback, 
causing M/L to rise toward
larger halos in proportion to the total-to-cold gas ratio. 
There seems to be enough energy in AGNs, though the actual feedback
mechanism is not properly understood yet.
\item
Protogalaxies below $M_{\rm shock}$ may be seen as Ly-$\alpha$ emitters
rather then as X-ray sources, while the hot gas in more massive halos
still produces X-rays.
\end{itemize}


\section{Photo-evaporation Scale}
\label{sec:evap}

Some of the halos below $V \sim 30\kms$ must be \emph{totally dark}, 
with no luminous component. 
Marginal evidence is starting 
to emerge from the statistics of image multiplicities
in gravitational lenses \cite{koch04}, 
and independently from the following comparison between theory and the observed
frequency of dwarf galaxies.  
While the faint-end galaxy luminosity function is roughly 
$\phi(L)\propto L^{-1}$ (or steeper), the halo mass function predicted in
$\Lambda$CDM is roughly $\psi(M)\propto M^{-2}$ (or flatter). These cannot be
reconciled with the virial relation $M \propto V^3$ and the Tully-Fisher
relation in the dwarf regime, $L \propto V^5$ (or flatter). 
If, however, only a fraction 
$f_{\rm L}(V)$ of the small halos have a luminous component
(with $L/M \propto V^2$ as discussed above),
then the observations can be reconciled once
$f_{\rm L} \prop V^2$.
Thus, if the dark-dark halos appear at $V\sim 30\kms$, we 
expect $\sim 90\%$ of the halos to be totally dark at the faint end, 
$V\sim 10\kms$. 

Supernova outflows alone are not likely to produce DDHs, both
because they must leave a trace of stars behind, and because they tend
not to be equally effective in all directions \cite{maclow99}.  
Ram pressure due to
outflows from nearby galaxies may remove gas in some cases \cite{scan00}, 
but probably not enough for the purpose.
On the other hand,
the origin of DDHs could be explained by \emph{radiative feedback}.
The first stars and AGNs produce UV flux which ionizes most of the gas in
the universe by $z_{\rm ion} \sim 10$, and keeps it at 
$T\sim (1-2)\times10^4$K until $z_{\rm end} \sim 2$ or so \cite{barkana01}. 
Further infall of gas is suppressed in halos below the Jeans scale,
which is crudely estimated to be about $V \sim 30\kms$ \cite{gnedin00},
but this does not explain the complete removal of the gas already
residing in those halos.
A significant fraction of $T\simeq 10^4$K gas is expected to escape
instantaneously from halos of a correspondingly shallower potential,
 $V < 10\kms$ \cite{barkana99},
but not from halos of $10<V<30\kms$.

\begin{figure}
\begin{center}
\includegraphics[width=.5\textwidth]
{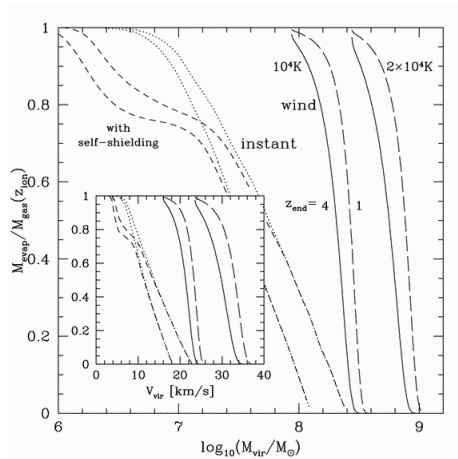}
\end{center}
\vspace{-10pt}
\caption[]{The evaporated mass fraction as a function of virial mass
and velocity (insert).
The curves on the left correspond to instantaneous expulsion,
and the curves on the right refer to steady evaporation, where
the ionization is assumed to be efficient from $z=8$ till either $z=4$ (solid)
or $z=1$ (dashed). The gas temperature is either $10^4$K (left) or
$2\times 10^4$K (right). While half the gas is lost instantaneously
from $V\sim 10\kms$ halos, a steady wind can remove a similar fraction
of the gas from $V\sim 30\kms$ halos.}
\label{fig:evap}
\end{figure}

However, if a continuous input of energy keeps the gas ionized for many
dynamical times, it can evaporate via a \emph{steady thermal wind} even
from deeper potential wells.
Once the gas is lost from the shallow edge of the potential well,
it is continuously replenished by interior gas, and so on. 
In Shaviv \& Dekel \cite{sd04}
we analyzed this process following the classic
solar-wind analysis \cite{parker60}, and found that a significant fraction
of the gas eventually evaporates from halos as large as $V \sim 30\kms$.
\fig{evap} summarizes the results of this analysis.

Thus, halos in the range $10<V<30\kms$ lose most of their gas after
the cosmological reionization. They may end up as gas-poor dwarf 
spheroidals with an old stellar population \cite{dw03} \cite{sd04}, or as DDHs
(see also \cite{som02}).

Halos of $V<10\kms$ are not expected to form stars at all because
of the cooling barrier at $T\sim 10^4$K (\fig{cool}). 
Atomic line cooling is very efficient just above this temperature,
but the only cooling agent below $10^4$K is molecular hydrogen, which 
cannot survive the UV background and is anyway very inefficient
\cite{haiman96}.

\section{Conclusion}
\label{sec:conc}

\begin{figure}[b]
\begin{center}
\includegraphics[width=.7\textwidth]
{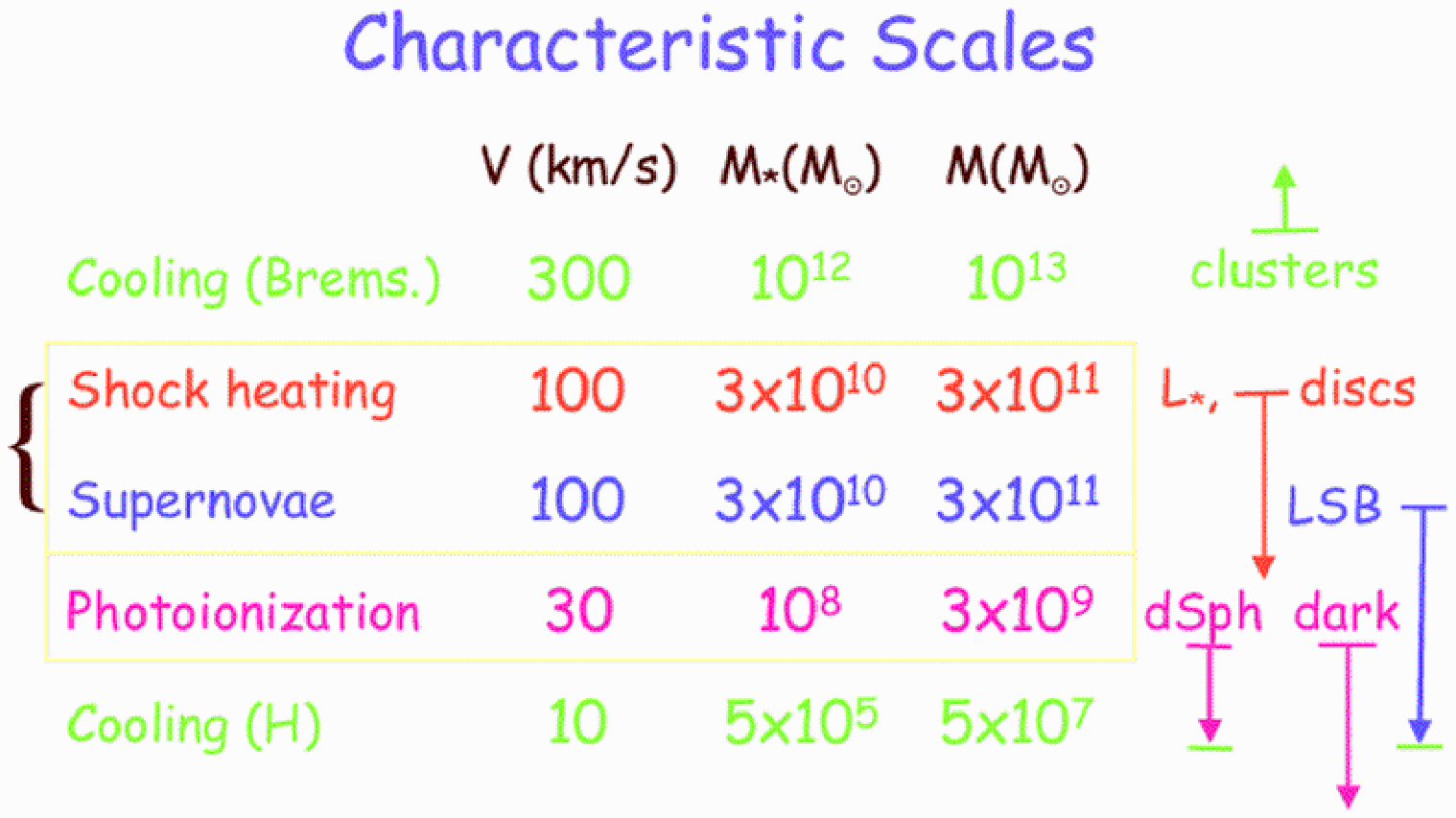}
\end{center}
\vspace{-25pt}
\label{fig:sum}
\end{figure}

The following table summarizes the characteristic scales originating from 
the physical processes discussed above and their possible association with the
observed scales. The classic argument of cooling on a vaguely-defined 
dynamical time provides relatively loose upper and lower bounds for luminous
galaxies.
The combination of (a) the newly discovered scale distinguishing between 
virial shock heating and cold flows and (b) 
the familiar supernova feedback scale 
seems to produce the robust bi-modality imprinted on the observed
galaxy properties, and in particular the characteristic 
upper limit for disk galaxies near $L_*$. 
This is provided that AGN feedback, or another process such as thermal
conductivity, indeed prevents the shock-heated gas in massive halos
from ever forming stars.  The details of the proposed scenario
are yet to be worked out (see also \cite{benson03} \cite{binney04}). 
While supernova feedback can be the primary process
responsible for the ``fundamental line" of LSB and dwarf galaxies, 
the radiative feedback due to cosmological reionization
may be responsible for totally evacuating small halos from gas
and producing dwarf spheroidals and dark-dark halos.
Once these physical process are properly incorporated in semi-analytic
models of galaxy formation, they may help solving the apparent conflicts
between theory and observation.

\medskip\noindent
Acknowledgment: I thank my collaborators Yuval Birinboimi and Nir Shaviv,
and those responsible for the 3D hydro simulations, R. Dave, N. Katz, 
D. Keres, A. Kravtsov, \& D. Weinberg.
This research has been supported by ISF 213/02 and NASA ATP NAG5-8218.


%

\end{document}

%

%
%
\section{Figures}
\begin{figure}[b]
\begin{center}
\includegraphics[width=.3\textwidth]{figure.eps}
\end{center}
\caption[]{Example of an electronically included eps-figure}
\label{eps1}
\end{figure}
%
%
%
%

\begin{table}
\caption{Critical $N$ values}
\begin{center}
\renewcommand{\arraystretch}{1.4}
\setlength\tabcolsep{5pt}
\begin{tabular}{llllll}
\hline\noalign{\smallskip}
${\mathrm M}_\odot$ & $\beta_{0}$ & $T_{\mathrm c6}$ & $\gamma$
  & $N_{\mathrm{crit}}^{\mathrm L}$
  & $N_{\mathrm{crit}}^{\mathrm{Te}}$\\
\noalign{\smallskip}
\hline
\noalign{\smallskip}
 30 & 0.82 & 38.4 & 35.7 & 154 & 320 \\
 60 & 0.67 & 42.1 & 34.7 & 138 & 340 \\
120 & 0.52 & 45.1 & 34.0 & 124 & 370 \\
\hline
\end{tabular}
\end{center}
\label{Tab1a}
\end{table}

\begin{table}
\caption{Please write your table caption here. Multi-line captions as
well as single-line captions automatically will be set flushleft}
\begin{center}
\renewcommand{\arraystretch}{1.4}
\setlength\tabcolsep{15pt}
\begin{tabular}{@{}llp{1.8cm}l}
\hline\noalign{\smallskip}
Nominal dimension & Angle & Tolerance & Evaluation factor \\
(mm) & & (\umu m) & \\
\noalign{\smallskip}
\hline
\noalign{\smallskip}
$10^1$ & $1^\circ$ & Untoleranced dimension& $c_1 =0$ \\
$10^{-1}$ & $1^{\prime\, {\mathrm a}}$ & 101--200 & $c_1 = 1$ \\
$10^{-2}$ & $1^{\prime\prime\, {\mathrm b}}$ & 40--100 & $c_1 = 2$ \\
$10^{-3}$ & & $< 50 $ & $c_1 = 3$ \\
\noalign{\smallskip}
\hline
\noalign{\smallskip}
\end{tabular}
\end{center}
$^{\mathrm a}$ One minute of arc. \\
$^{\mathrm b}$ One second of arc.
\label{Tab1b}
\end{table}

%